\documentclass[prc,superscriptaddress,showpacs,twocolumn]{revtex4}
\usepackage{bm}% bold math
\usepackage{dcolumn}
\usepackage{graphicx}% Include figure files
\usepackage{amsmath}
\usepackage{amssymb}
\usepackage{color}
\newcommand{\lm}{\Lambda}
\newcommand{\lb}{\Lambda_{\rm b}}
\newcommand{\be}{\begin{equation}}
\newcommand{\ee}{\end{equation}}
\newcommand{\vnn}{V_{\rm NN}}
\newcommand{\vlowk}{V_{{\rm low}\,k}}
\newcommand{\fmi}{\, \text{fm}^{-1}}
\newcommand{\mev}{\, \text{MeV}}

\newcommand{\beqa}{\begin{eqnarray}}
\newcommand{\eeqa}{\end{eqnarray}}

\begin{document}
\title{Matter and charge radius of $^6$He in the hyperspherical-harmonics approach}

\author{S.\ Bacca}
\email[E-mail:~]{bacca@triumf.ca}
\affiliation{TRIUMF, 4004 Wesbrook Mall, Vancouver, BC, V6T 2A3, Canada}
\author{N.\ Barnea}
\email[E-mail:~]{nir@phys.huji.ac.il}
\affiliation{Racah Institute of Physics, Hebrew University, 91904, Jerusalem,
Israel}
\author{A.\ Schwenk}
\email[E-mail:~]{schwenk@physik.tu-darmstadt.de}
\affiliation{ExtreMe Matter Institute EMMI, 
GSI Helmholtzzentrum f\"{u}r Schwerionenforschung GmbH, 
64291 Darmstadt, Germany}
\affiliation{Institut f\"{u}r Kernphysik, 
Technische Universit\"{a}t Darmstadt, 64289 Darmstadt, Germany}

\begin{abstract}
We present {\it ab-initio} calculations of the binding energy and
radii of the two-neutron halo nucleus $^6$He using two-body
low-momentum interactions based on chiral effective field theory
potentials. Calculations are performed via a hyperspherical
harmonics expansion where the convergence is sped up introducing an
effective interaction for non-local potentials. The latter is
essential to reach a satisfactory convergence of the extended matter
radius and of the point-proton radius.  The dependence of the
results on the resolution scale is studied.  A correlation is found
between the radii and the two-neutron separation energy.  The
importance of three-nucleon forces is pointed out comparing our
results and previous calculations to experiment.
\end{abstract}

\pacs{21.10.Dr, 21.10.Gv, 21.60.De, 27.20.+n
%21.10.Dr Binding energies and masses
%21.10.Gv Nucleon Distributions and halo features
%21.60.De Ab-initio methods
%27.20.+n 6<=A<=19
}
\maketitle

\section{Introduction}

The physics of strong interactions gives rise to fascinating phenomena
like the formation of a halo structure where one or more loosely-bound
nucleons surround a tightly bound core. The lightest known halo system
is the $^6$He nucleus, made of two neutrons and a $^4$He
core~\cite{Tan96}.  This nucleus is of Borromean nature, where the
two-neutron and neutron-core subsystems are unbound, but the
three-body system is held together~\cite{Jon04}.  $^6$He is a
radioactive nucleus that undergoes $\beta$-decay with a half-life of
0.8 s~\cite{Knecht}.  Despite this short life-time, a combination of
nuclear and atomic physics techniques recently enabled a series of
precision measurements of $^6$He.  The ground-state energy has been
measured directly for the first time at TITAN~\cite{Maxime}, leading
to an improved value of its charge radius~\cite{Wan04,Mue07}.
Understanding and predicting these extreme phenomena presents a great
testing ground for theory, leading to a deeper understanding of the
strong force and nuclear interactions for neutron-rich systems.  As a
light nucleus, the study of $^6$He is amenable to {\it ab-initio}
methods starting from nuclear forces.  The simultaneous description of
the large radius and small binding of the halo neutrons makes the
reproduction of the precise experimental data for halo nuclei
particularly challenging.

One of the central advances in nuclear theory has been the development
of chiral effective field theory (EFT) for nuclear
forces~\cite{EHM,EM}, rooted in quantum cromodynamics (QCD). In this
formalism nucleon-nucleon (NN) interactions and many-body forces are
constructed systematically and consistently.  Before that, the
traditional models to describe the interaction among nucleons were
based on phenomenology or meson exchange theories.  For light nuclei,
{\it ab-initio} methods have established the quantitative importance
of three-nucleon (3N) forces for ground-state properties, excitations
and
reactions~\cite{GFMC_Enrico_Fermi_School,Nollett,Navratil09,Bacca_PRL09}.
In addition, first approximate studies have shown new facets of 3N
forces in heavier neutron-rich
nuclei~\cite{achim1,achim2,achim3}. Despite these developments, 3N
forces based on chiral EFT remain unexplored in halo nuclei.

In the literature, several {\it ab-initio} calculations of halo nuclei
with traditional nucleon-nucleon (NN) potentials exist for both energy
and radii, e.g., within the No-Core Shell Model (NCSM) with
meson-exchange and phenomenological interactions~\cite{NCSMHe},
Fermionic Molecular Dynamics (FMD) studies based on the Unitary
Correlation Operator Method (UCOM) potential~\cite{FMD}, and the
Microscopic Cluster Model (MCM) with semi-realistic
interactions~\cite{Brida}.  The only existing converged calculations
that explicitly include 3N forces are based on the Green's Function
Monte Carlo (GFMC) method~\cite{GFMC_Enrico_Fermi_School}.  While
light nuclei have been investigated with chiral EFT potentials using
the NCSM~\cite{Navratil09}, there are no results for $^6$He based on
chiral NN and 3N interactions. The rapid Gaussian fall-off of the
NCSM wave function does not make it the optimal method to investigate
the extended halo nuclei. The correct description of the exponential
fall-off can be achieved by combining the NCSM with the Resonating
Group Method, as done in~\cite{Sofia} for one-neutron halo nuclei.

In this paper, we present a chiral EFT based study of $^6$He, limited
to two-body forces, as a first step towards predicting halo nuclei from EFT.
Some of the results of this work have been
presented in~\cite{Maxime}. Here we give a complete overview of our
theoretical study and explain the calculations in detail.

We combine the renormalization group (RG) evolution of chiral EFT
potentials to low-momentum interactions~\cite{Vlowk} with the {\it
  ab-initio} Hyperspherical Harmonics (HH) method for
$^6$He~\cite{Nir}.  Our work goes beyond the previous coupled-cluster
theory investigation~\cite{Hag06} of the helium isotopes in that we
study the cutoff variation, as a tool to probe the effects of
many-body forces on the binding energy and also on the radii of
$^6$He.  A similar study of the cutoff variation was performed
in~\cite{SRG_NCSM} using similarity-RG-evolved EFT potentials within
the NCSM, where only binding energies were investigated within a
limited Hilbert space ($N_{\rm max}=10$) for $^6$He.  As opposed to a
shell model expansion, the use of the hyperspherical basis enables 
to better describe the exponential fall-off of the nuclear wave
function, which is important to provide precise results for  $^6$He. 
This work also
goes beyond our first study~\cite{Sonia_EPJ}, since here we present
converged results for the energy and new calculations of radii for
$^6$He, where the use of the effective interaction method enables to
speed up the convergence.

This paper is organized as follows. In Section II we describe the
theoretical aspects of the interaction and of the few-body method. In
Section III we present our results for the binding energy and radii of
$^6$He, which are compared to the experimental data in Section
IV. Finally, we conclude in Section V.

\section{Theoretical aspects}

\subsection{Nuclear forces}

Nuclear forces depend on a resolution scale, which we denote by a
generic momentum cutoff $\Lambda$, and the Hamiltonian consists of NN
and corresponding many-body (3N, 4N, ....)  interactions~\cite{EHM,EM}:
\be
H(\lm) = T + \vnn(\lm) + V_{\rm 3N}(\lm) + V_{\rm 4N}(\lm) + \ldots \,.
\label{Hamiltonian}
\ee
For most nuclei, the typical momenta are of the order of the pion
mass, $Q \sim m_\pi$, and therefore pion exchanges are included
explicitly in nuclear forces. In chiral EFT~\cite{EHM, EM}, nuclear
interactions are organized in a systematic expansion in powers of
$Q/\lb$, where $\lb$ denotes the breakdown scale, roughly $\lb \sim
m_\rho$. At a given order, this includes contributions from one- or
multi-pion exchanges and from contact interactions, with short-range
couplings that depend on the resolution scale $\lm$ and for each $\lm$
are fit to data. At N$^3$LO [or $(Q/\Lambda_b)^4$], chiral NN
interactions accurately reproduce low-energy NN scattering~\cite{EHM,EM}.

In the current study we take the Entem and Machleidt N$^3$LO chiral NN
potential with $\Lambda = 500$ MeV~\cite{EM_pot} as a starting point,
and then use the RG~\cite{Vlowk} to evolve it to low-momentum
interactions $\vlowk$ with $\Lambda=1.8$--$2.4 \fmi$ ($360$--$480
\mev$).  To the evolved Hamiltonian, we add the Coulomb and other
standard electromagnetic interactions.  The RG preserves the
long-range pion exchanges and includes subleading contact
interactions, so that NN scattering data are reproduced.  We use the
RG to soften the short-range repulsion and short-range tensor
components of the initial chiral interactions so that convergence of
the few-body calculations is vastly accelerated.  The cutoff variation
of few- and many-body observables then serves as an estimate of the
theoretical uncertainty due to neglected 3N and higher-body forces.
The addition of evolved 3N forces, which is expected to reduce the cutoff dependence,
will be pursued in future work using the Similarity RG evolution in momentum
space \cite{Hebeler}.
\subsection{Hyperspherical-harmonics approach}

Given the Hamiltonian $H$ we use the HH expansion to solve the
Schr\"{o}dinger equation.  The HH method, typically employed in
few-body physics to study nuclei with mass number A=3 and 4, can be
extended to the investigation of $^6$He (see,
e.g.,~\cite{He6Sonia1,He6Sonia2}), using a powerful antisymmetrization
algorithm introduced in~\cite{Nir}.  This approach is translationally
invariant, being constructed with the Jacobi coordinates
\begin{eqnarray}
\label{jacobc}
\boldsymbol {\eta }_{0~~}&=&\frac{1}{\sqrt{A}}\sum _{i=1}^{A}\mathbf{r}_{i}\,,\\
\nonumber
\boldsymbol {\eta }_{k-1}&=&\sqrt{\frac{k-1}{k}}\left( \mathbf{r}_{k}-\frac{1}{k-1}\sum _{i=1}^{k-1}\mathbf{r}_{i}\right),\, k=2,...,A\,,
\end{eqnarray}
where $\mathbf{r}_i$ are the particle coordinates.  Using the
$\boldsymbol {\eta }_{i}$ one can construct the hyperspherical
coordinates composed of one hyperradial coordinate
$\rho=\sqrt{\sum_{i=1}^{A-1} \boldsymbol {\eta }_{i}^2}$ and a set of
$(3A-4)$ angles that we denote with $\Omega$ (for more details
see~\cite{Nir}).
%The HH expansion method possesses by construction
%the correct exponential asymptotic behavior, making it an optimal
%method to study halo nuclei. 
In the HH method, the wave-function expansion reads
\begin{eqnarray}
\label{expans}
\lefteqn{\Psi( \boldsymbol {\eta }_{1}, ..., \boldsymbol {\eta }_{A-1},
  s_1,...,s_A, t_1,...,t_A)= } \\
\nonumber
& & \sum_{[K] n}^{K_{\rm max}, n_{\rm max}}C_{[K] n} \, R_{n}(\rho) \,
{\cal Y}_{[K]}(\Omega,s_1,...,s_A, t_1, ...,t_A ),
\end{eqnarray}
where $s_i$ and $t_i$ are the spin and isospin of nucleon i,
respectively, $C_{[K] n}$ is the coefficient of the expansion, labeled
by $[K]$, which represents a cumulative quantum number that includes
the grandangular momentum $K$ related to the hyperspherical harmonics
${\cal Y}_{[K]}$, and $n$ labels the hyperradial wave function
$R_{n}(\rho)$.  The latter is given by
\begin{equation}
R_{n}(\rho) =\sqrt{\frac{n!}{(n+a)!}} b^{\frac{-3(A-1)}{2}}
\left(\frac{\rho}{b}\right)^{\frac{a-(3A-4)}{2}}
 L^a_{n}\left(\frac{\rho}{b}\right)
 e^{-{\rho}/{2b}},
\label{laguerre}
\end{equation}
where $L^a_{n}$ are the generalized Laguerre polynomials,  and $b$ is a
scale parameter. Equation (\ref{laguerre}) shows that the hyperradial
basis functions fall-off exponentially as $-b/2\rho$.  The Laguerre
polynomials provide a power expansion in $\rho/b$ which gives sufficient flexibility
to describe the wave function in the short and intermediate range.
As the wave-function expansion contains more and more hyperadial terms,
i.e., Laguerre functions in Eq.~(\ref{laguerre}), the correct hyperradial structure
of the nuclear wave function is recovered regardless of the value of $b$.
With about 40 hyperradial states, we observed a 0.05$\%$ relative 
change in the energy when b is varied by one order of magnitude.

Recently, we extended the use of the HH basis for non-local
interactions~\cite{HHJISP,HHUCOM}, by expanding the potential in
harmonic-oscillator matrix elements.  In~\cite{Sonia_EPJ} we used a HH
expansion to study the energy of $^6$He with $\vlowk$ potentials. A
rather slow convergence was observed, especially for the larger values
of $\Lambda$, and we therefore performed an exponential extrapolation
of the variational HH energies to the infinite Hilbert space.  The
same variational argument, however, cannot be used for other
observables, like the radii. To further speed up the convergence of
the calculation, we employed an effective interaction of the
Lee-Suzuki type.  This method, called Effective Interaction
Hyperspherical Harmonics (EIHH), was first introduced
in~\cite{EIHHlocal} and recently extended to non-local potentials
in~\cite{EIHHnonlocal}.  In the following, we present our results for
the ground-state energies and radii which we obtained with the EIHH
method using chiral low-momentum potentials.

\section{Results and Discussion}

The main focus of this work is to study the $^6$He
nucleus. Nevertheless, we first consider $^4$He. This allows us to
highlight the different convergence patterns between the HH and EIHH
methods, as well as to benchmark our results with those of other
few-body techniques.

\begin{figure}
\includegraphics[scale=0.35,clip=]{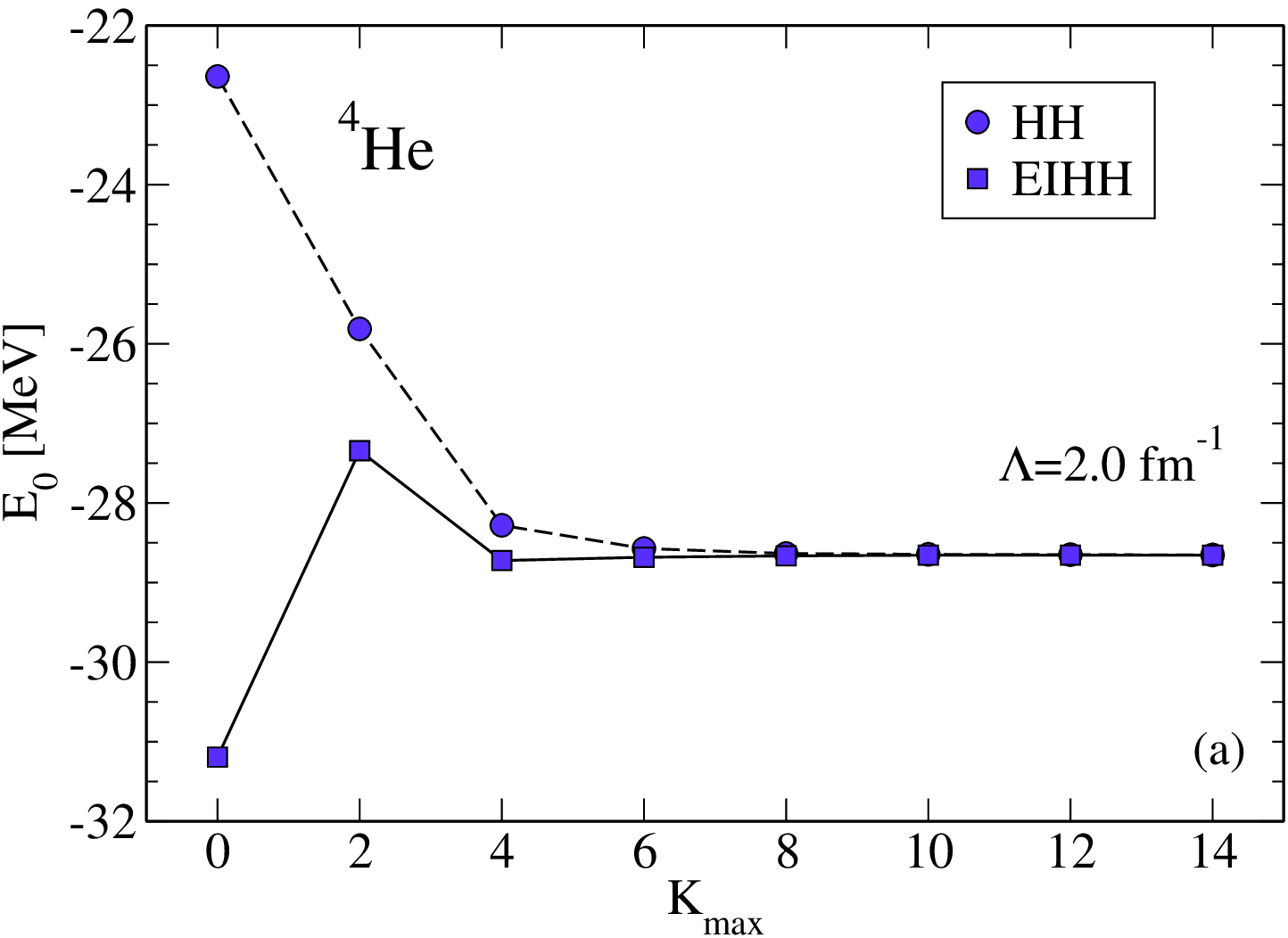}
\includegraphics[scale=0.35,clip=]{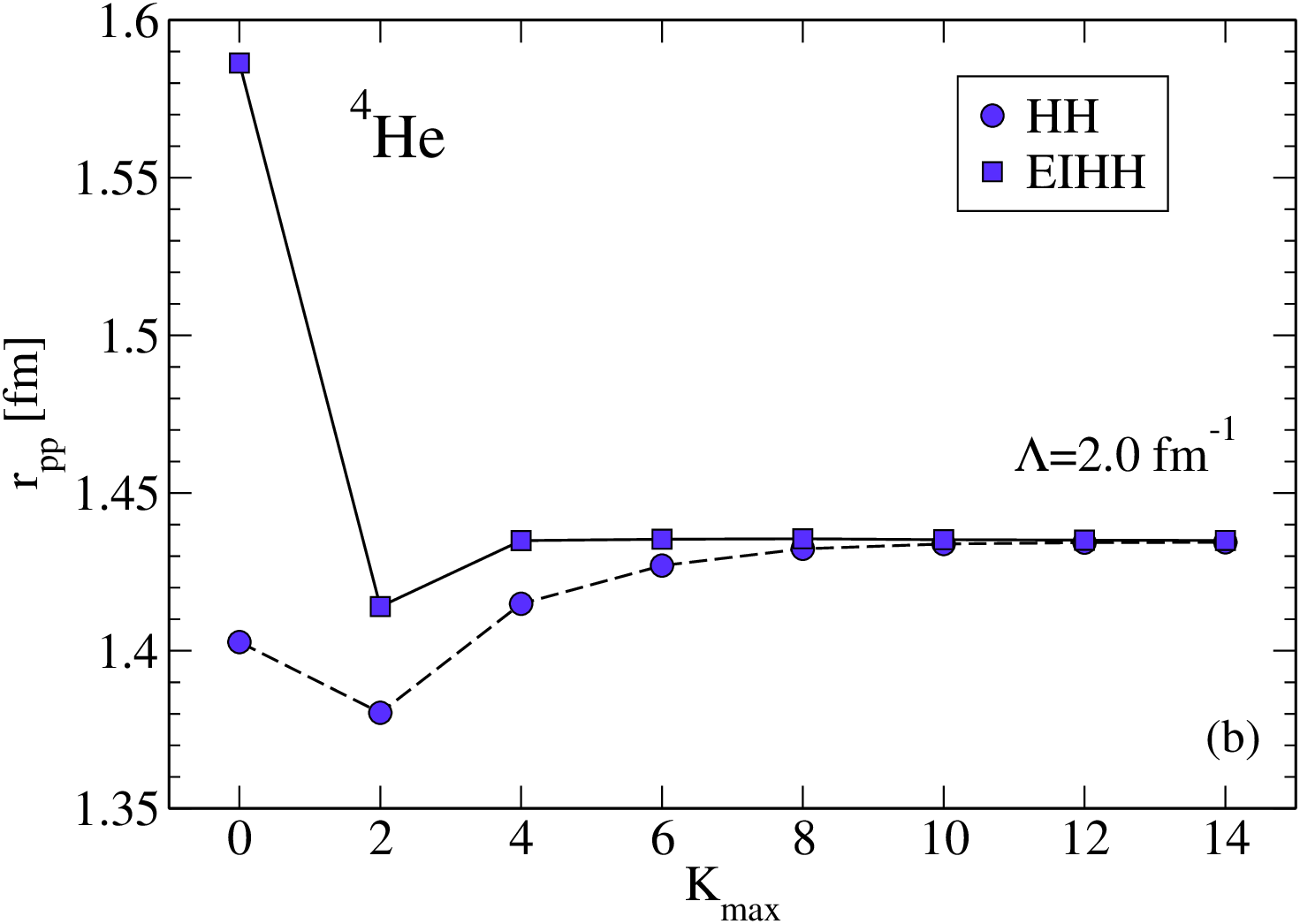}
\caption{(Color online) $^4$He ground-state energy (upper panel) and
  rms matter radius (lower panel) as a function of the maximum
  grandangular momentum $K_{\rm max}$ used in the wave-function
  expansion. The convergence pattern of the HH (circles) expansion is
  compared to the EIHH (squares).  The cutoff of the $\vlowk$ chiral
  potential is $\Lambda=$2.0 fm$^{-1}$.}
\label{he4_energy_radius}
\end{figure}

In Fig.~\ref{he4_energy_radius}, we compare the convergence pattern of
the HH to the EIHH approach for $^4$He.  The ground-state energy and
rms (root-mean-square) matter radius are presented as a function of
the maximum grandangular momentum $K_{\rm max}$ used in the
wave-function expansion. Larger spaces are accessible for $^4$He but
are not necessary for low-momentum interactions.  The cutoff of the
$\vlowk$ chiral potential used in Fig.~\ref{he4_energy_radius} is
$\Lambda=$2.0 fm$^{-1}$, but a very similar convergence is obtained
for cutoffs in the range $1.6 \leqslant \Lambda \leqslant 2.8$
fm$^{-1}$.  The energy is in agreement with the Faddeev-Yakubovsky
result of -28.65(5) MeV~\cite{Nogga} with the same potential.
For the cutoff range $\Lambda = 1.8 - 2.4 \fmi$, the $^4$He ground-state
energy varies from $29.30 - 27.40 \mev$ [24]. In addition, we remind the
reader that if one uses the Entem and Machleidt N$^3$LO chiral NN
potential~\cite{EM_pot}, without evolving it to $\vlowk$, the $^4$He
ground-state energy is $-25.38 \mev$~\cite{EIHHnonlocal}.
 The
radius obtained with the EIHH has already been benchmarked with other
few-body methods in~\cite{EIHHnonlocal} using the N$^3$LO chiral
potential~\cite{EM_pot}.   From Fig.~\ref{he4_energy_radius} it is
apparent that, even though an excellent convergence can be reached
within the model space for both the HH and EIHH case, the EIHH
convergence rate is superior, since with $K_{\rm max}=4$ one has
already reached the converged value at a subpercentage level.  We also
note that in case of the HH expansion, the radius converges slower
than the energy. In fact, at $K_{\rm max}=8$ the radius is converged
only within 0.15$\%$ whereas the energy is within 0.06$\%$. In the
same $K_{\rm max}$ space, the EIHH results are converged to 0.03$\%$
both for the energy and the radius.

\begin{figure}
\includegraphics[scale=0.45,clip=]{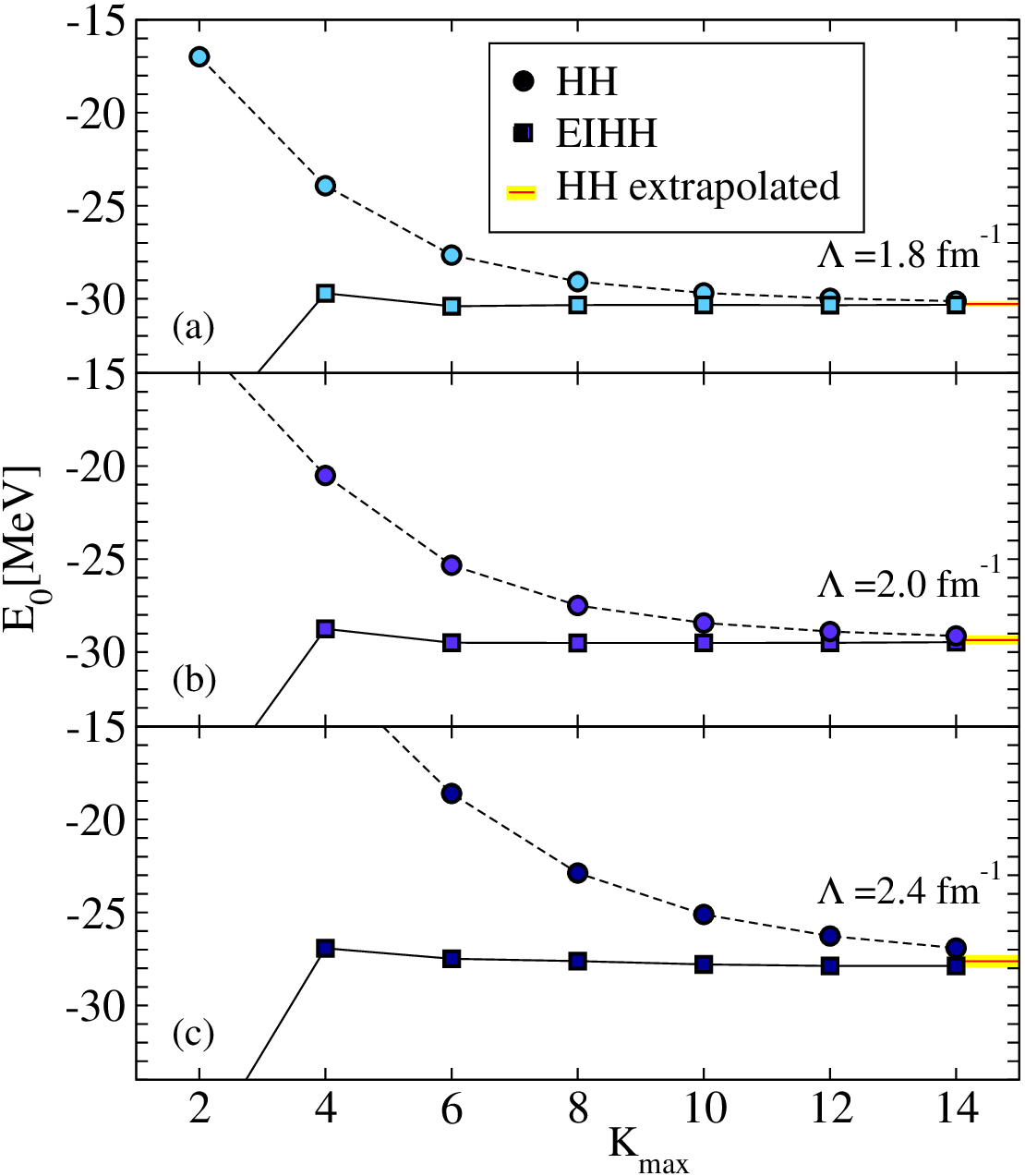}
\caption{(Color online) $^6$He ground-state energies as a function of
  the grandangular momentum $K_{\rm max}$ obtained for three different
  values of the cutoff $\Lambda=1.8,2.0,2.4$ fm$^{-1}$ of the $\vlowk$
  chiral potential. The HH convergence pattern is taken
  from~\cite{Sonia_EPJ} and plotted together with the new EIHH
  results. The value obtained from the exponential extrapolation of
  the HH results with the corresponding error bar
  (see~\cite{Sonia_EPJ}) is also shown.}
\label{he6_energy}
\end{figure}

\begin{figure}
\includegraphics[scale=0.45,clip=]{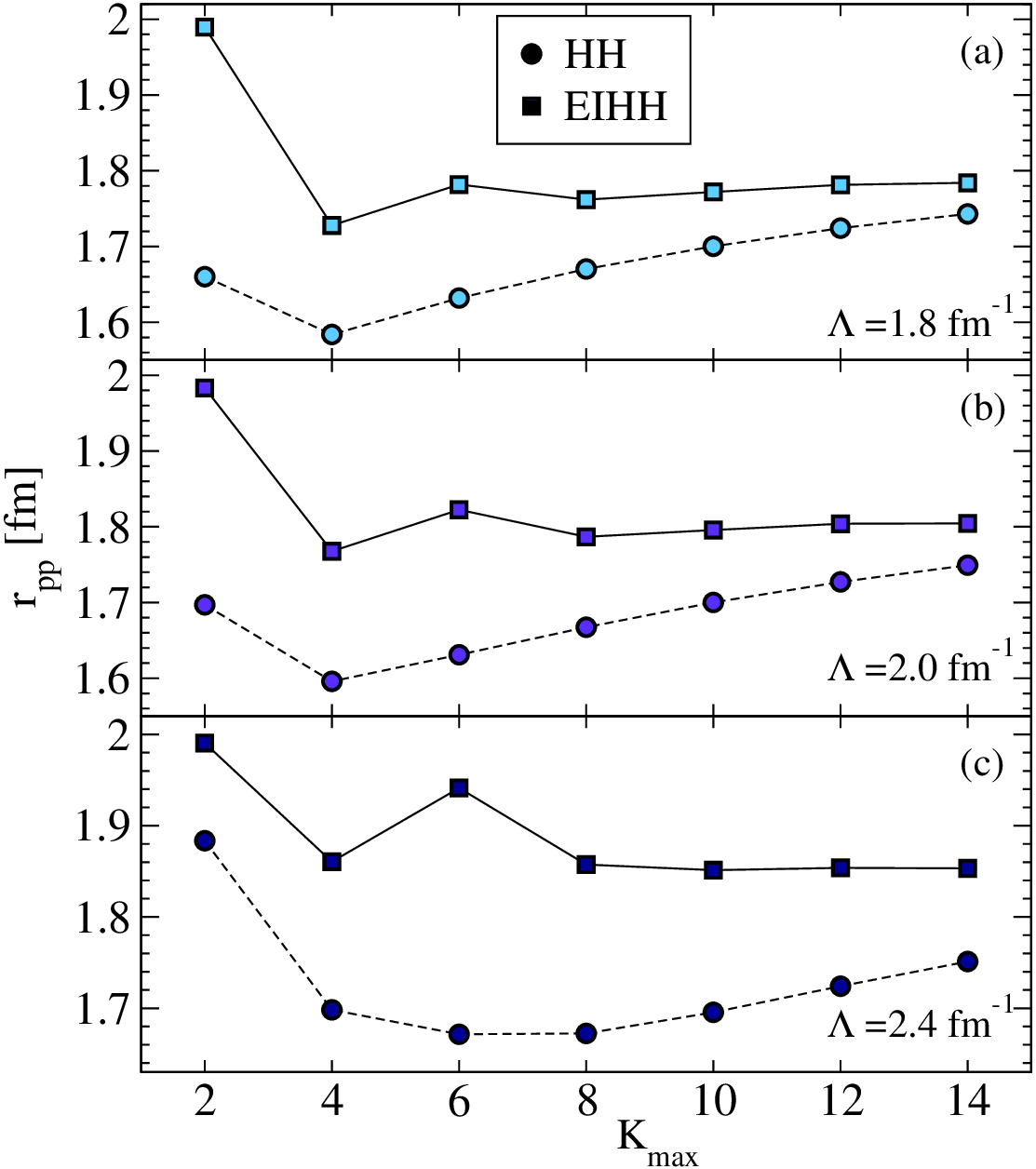}
\caption{(Color online) $^6$He rms point-proton radius as a function
  of the grandangular momentum $K_{\rm max}$ obtained for three
  different values of the cutoff $\Lambda=1.8,2.0,2.4$ fm$^{-1}$ of
  the $\vlowk$ chiral potential. The HH and EIHH convergence patterns
  are compared.}
\label{he6_radius}
\end{figure}

We can now focus on the $^6$He nucleus.  In Fig.~\ref{he6_energy}, we
present the convergence patterns of the HH and EIHH methods for the
ground-state energy as a function of the grandangular momentum $K_{\rm
  max}$.  The patterns are presented for three different cutoffs
$\Lambda=1.8,2.0,2.4$ fm$^{-1}$ of the $\vlowk$ chiral potential. 
 The
maximum value of $K_{\rm max}$ that we can reach with the present
codes and computing capability for $^6$He is 14, which corresponds to
about $3\times 10^6$ basis states.  
The cutoff range $\Lambda = 1.8 - 2.4 \fmi$ has been chosen to speed up
the convergence of the basis expansion. When $\Lambda$ increases, the
result of the starting Entem and Machleidt potential~\cite{EM_pot} should be
recovered. Converged results, especially for the radii, are however very
hard to achieve within a $K_{\rm max}=14$ model space.

The HH results are taken
from~\cite{Sonia_EPJ} and plotted together with the new EIHH
results. The value obtained from the exponential extrapolation of the
HH results with error bar obtained as explained in~\cite{Sonia_EPJ} is
also shown.  One can readily see that the convergence is faster when
the lowest cutoff $\Lambda=1.8$ fm$^{-1}$ is used, because the
interaction is softer. In this case, in fact, the curve for the HH
expansion reaches the EIHH one within the model space size of $K_{\rm
  max}=14$. For the larger cutoff values, and most evident for
$\Lambda=2.4$ fm$^{-1}$ this is not the case, because the HH
convergence is too slow to allow a satisfactory convergence within the
accessible spaces. In~\cite{Sonia_EPJ} we extrapolated the variational
HH results. It is reassuring that the HH extrapolated results agree
with the EIHH energies within the error bar for all the cutoffs used.
Even though an extrapolation of the variational results is justifiable
for ground-state energies, the use of the EIHH approach enables one to
avoid extrapolations. One can readily see in Fig.~\ref{he6_energy}
that the convergence pattern of the EIHH method is excellent, as
proven by the nice flattening of the curve from $K_{\rm max}=4$. For
this $K_{\rm max}$ space size and for the larger cutoff $\Lambda=2.4$
fm$^{-1}$ the HH result is still about 15 MeV above the converged
value, whereas the EIHH is within 3$\%$ of it.

We can now turn to the calculation of the radii. In case of a halo
nucleus it is interesting to study both the rms matter radius and the
rms point-proton radius or rms point-neutron radius.  They are
calculated as the expectation values of the operator $r_{\rm
  m/pp/pn}=\sqrt{\langle \psi_0 \left| r^2_{\rm m/pp/pn}\right|
  \psi_0\rangle}$, where
\begin{eqnarray}
r^2_{\rm m}&=&\frac{1}{A}\sum_i^A( \boldsymbol{r}_i- \boldsymbol{R}_{\rm cm})^2 \,, \cr
r^2_{\rm pp}&=&\frac{1}{Z}\sum_i^A( \boldsymbol{r}_i - \boldsymbol{R}_{\rm cm})^2 
\left(\frac{1 + \tau_i^3}{2}\right)\,, \cr
r^2_{\rm pn}&=&\frac{1}{N}\sum_i^A( \boldsymbol{r}_i - \boldsymbol{R}_{\rm cm})^2 
\left(\frac{1 - \tau_i^3}{2}\right)\,,
\label{radii_eq}
\end{eqnarray}
are the matter radius, the point-proton, and the point-neutron radius
operators, respectively.  Here ${\boldsymbol R}_{\rm cm}=\frac{1}{\sqrt{A}}\,
\boldsymbol{\eta}_0$ is the center-of-mass coordinate and $\tau^3_i$
is the third component of the isospin of nucleon $i$.

In Fig.~\ref{he6_radius}, the convergence patterns of the HH and EIHH
methods are shown for the case of the point-proton radius as a
function of $K_{\rm max}$. The three different values of the cutoff
$\Lambda=1.8,2.0,2.4$ fm$^{-1}$ of the $\vlowk$ chiral potential have
been used.  As already observed for $^4$He, one can see that the
convergence of the HH method is slow for the radius. For $^6$He it is
so slow that convergence cannot be reached within the accessible
$K_{\rm max}$ spaces.  With the EIHH method the convergence pattern  is
improved significantly and allows to provide solid result for this
observable.

\begin{figure}
\includegraphics[scale=0.35,clip=]{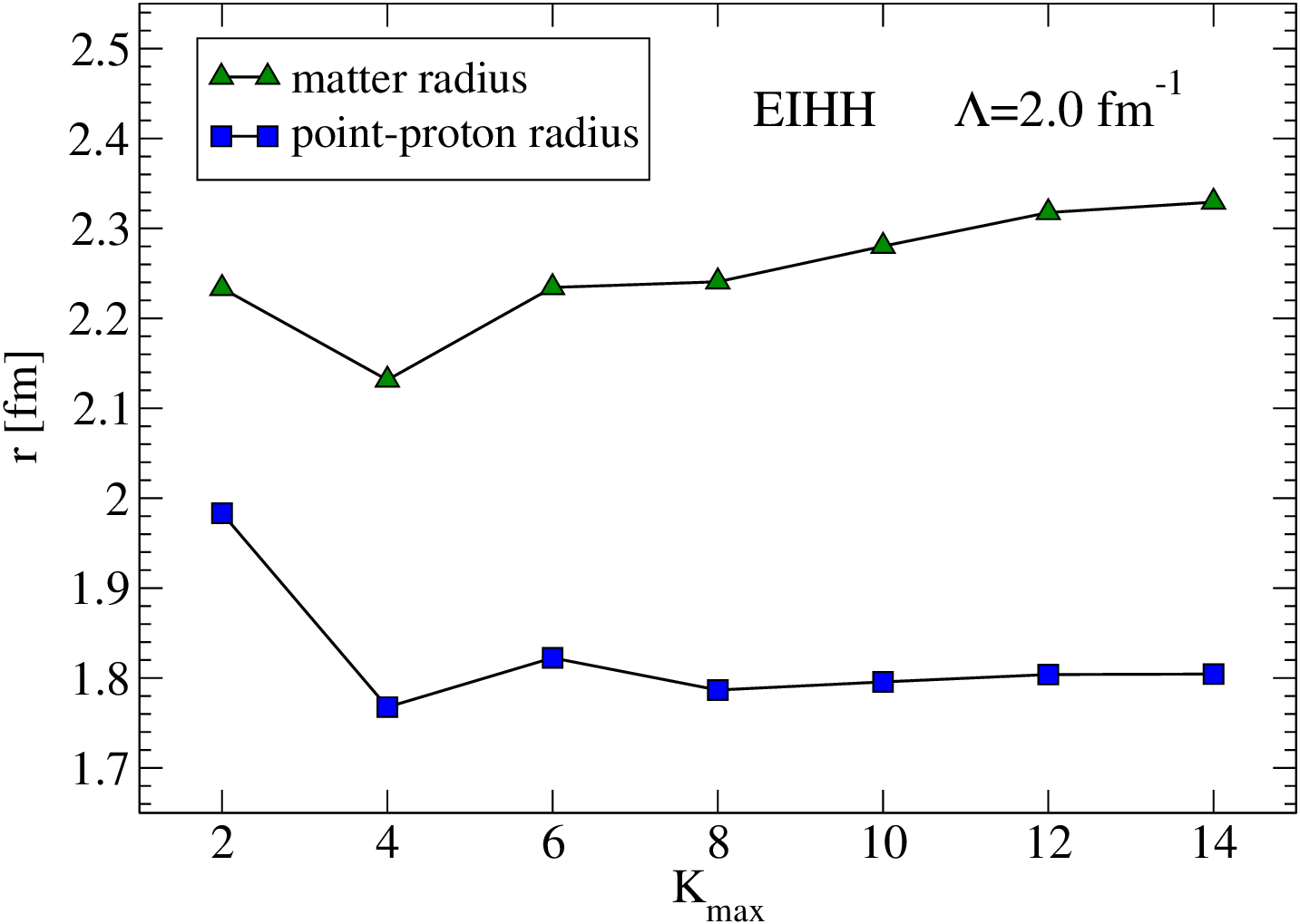}
\caption{(Color online) EIHH results of the $^6$He matter radius and
  point-proton radius as a function of the grandangular momentum
  $K_{\rm max}$ obtained for a cutoff value $\Lambda=2.0$ fm$^{-1}$ of
  the $\vlowk$ chiral potential.}
\label{he6_rm_rpp}
\end{figure}

In Fig.~\ref{he6_rm_rpp}, we compare the convergence of the matter and
point-proton radii for $^6$He for a cutoff of $\Lambda=$2.0 fm$^{-1}$.
The matter radius is about 30$\%$ larger than the point-proton radius
due to the neutron halo. This is in agreement with the finding of
other calculations, like the NCSM~\cite{NCSMHe}.  As a
result of the halo, we also observe that the convergence of the matter
radius is slower than that of the point-proton radius.  In order to
accurately describe the outer two halo neutrons, which contribute to
the matter radius but not directly to the point-proton radius [except
for a simple center-of-mass shift, see Eq.~(\ref{radii_eq})], a larger
space is required.

\begin{table}
\caption{$^4$He point-proton radius and $^6$He point-proton,
  point-neutron and matter radius for three different cutoffs
  $\Lambda=1.8,2.0,2.4$ fm$^{-1}$  obtained with the EIHH method. All
  radii are in fm. The two-neutron separation energy $S_{2n}$ is shown as well in MeV.}
\begin{center}
\begin{tabular}{c|c|c|c|c|c}
\hline\hline
$\Lambda$ [fm$^{-1}$]&  $^4$He $r_{\rm pp}$  &  $^6$He $r_{\rm pp}$  &  $^6$He $r_{\rm pn}$ &  $^6$He $r_{\rm m}$ & $^6$He $S_{2n}$ \\
\hline
1.8 &1.427(7)&1.78(1)&2.51(6)&2.30(6)&1.036(7)\\
2.0 &1.435(7)&1.804(9)&2.54(6)&2.33(5)&0.82(4)\\
2.4 &1.464(7)&1.853(9)&2.60(4)&2.39(3)&0.48(9)\\
\hline\hline
\end{tabular}
\end{center}
\label{table_radii}
\end{table}

In Table \ref{table_radii}, we present our EIHH results for the
point-proton, matter and point-neutron radii of $^6$He for different
cutoffs and compare it to the $^4$He radii as a reference. As an
estimate of the theoretical error associated with the few-body method
we take the difference between the largest possible calculation
($K_{\rm max}=14$) and the $K_{\rm max}=10$ result, or minimally
0.5$\%$.

As mentioned before the errors are larger for the matter and
point-neutron radii than for the point-proton radius.  For all three
cutoffs studied, we find that the $^6$He point-proton radius is larger
than the $^4$He point-proton radius by about 25-27$\%$. Similar
enhancements have been found within the NCSM~\cite{NCSMHe} using
different two-body potentials. The $^6$He $r_{\rm pp}$ is larger than
the $^4$He $r_{\rm pp}$ because of two effects: recoil of the
center-of-mass due to the presence of the halo neutrons in $^6$He and
core polarization. In our approach these two effects
are both taken into account and cannot be separated.  As shown in
Table \ref{table_radii}, we find that the point-neutron radius is
about $40\%$ larger than the point-proton radius, similar to other
 calculations~\cite{NCSMHe,Brida}.

Next, we discuss the cutoff dependence of our results with two-body
interactions only.  The cutoff variation of the energy and radii is
significantly larger than the theoretical uncertainties due to the
few-body method, which we have estimated above. The cutoff variation
is due to neglected three-body and many-body interactions in the
Hamiltonian $H(\lm)$. Varying the cutoff over the range $1.8-2.4$
fm$^{-1}$ leads to a running of about $8\%$ in the energy and of $4\%$
in the radii. The latter are long-range observables, which are less
sensitive to the $\Lambda$-dependence.  Our results clearly highlight the
fact that 3N forces are required for energies and radii.

\begin{figure}
\includegraphics[scale=0.35,clip=]{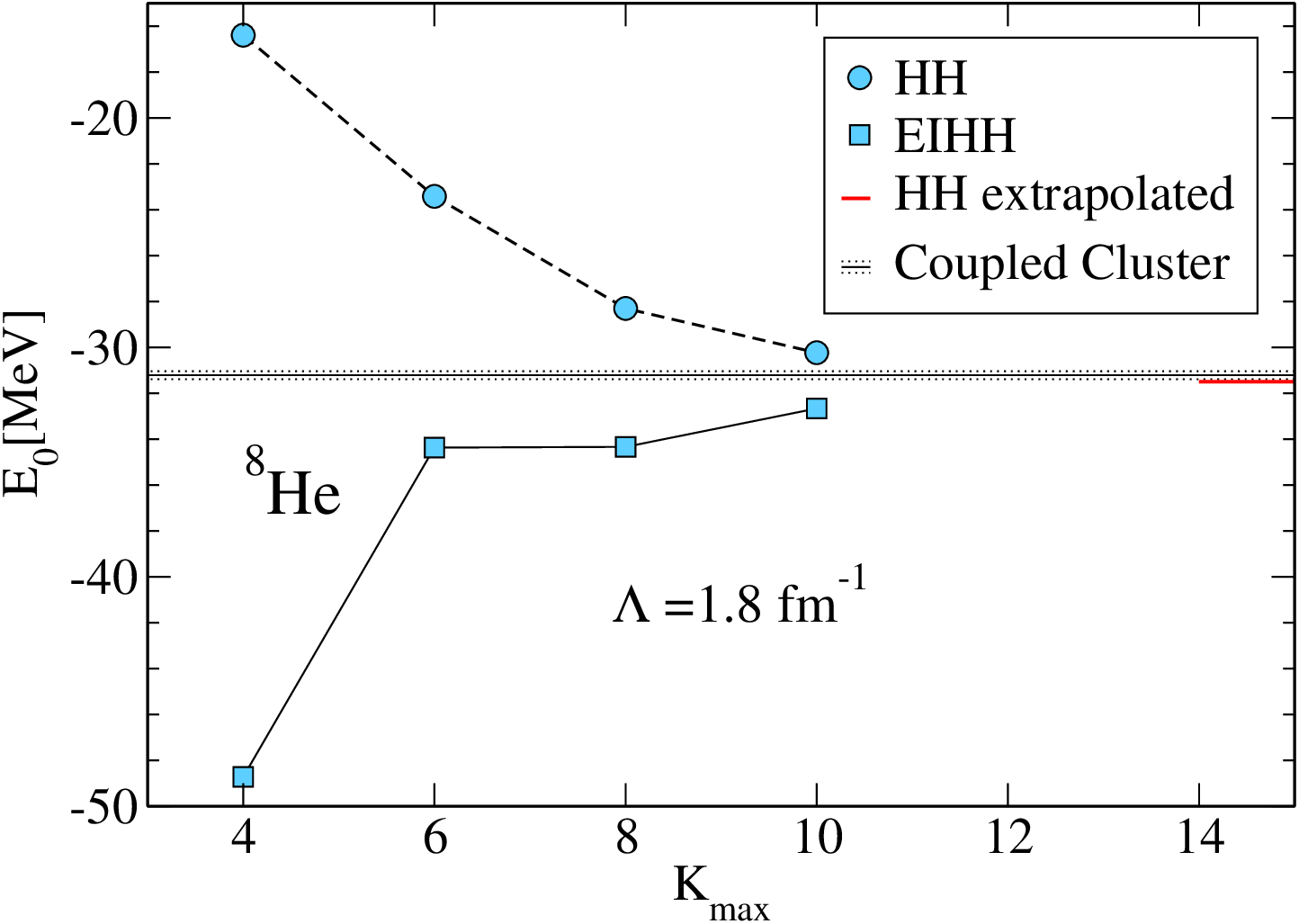}
\caption{(Color online) $^8$He ground-state energy calculated with the
  EIHH and HH method as a function of the grandangular momentum
  $K_{\rm max}$ obtained for a cutoff of $\Lambda=1.8$ fm$^{-1}$ of
  the $\vlowk$ chiral potential. The extrapolated HH results are shown
  as a reference and compared to the coupled cluster results
  from~\cite{Sonia_EPJ}.  }
\label{he8}
\end{figure}

We have also explored the four-neutron halo nucleus $^8$He with the HH
expansion.  In Fig.~\ref{he8}, we show the $^8$He ground-state energy
from a $\vlowk$ chiral potential with $\Lambda=1.8$
fm$^{-1}$. Convergence patterns are shown for the HH and EIHH methods
as a function of the grandangular momentum $K_{\rm max }$. We observe
that the convergence of the calculations is very slow. With the
present codes and computing capabilities we are not able to access
larger $K_{\rm max}$ spaces, so that fully converged calculations
cannot be provided at the moment.  We have tried an exponential
extrapolation of the HH results and present it as a reference in
Fig.~\ref{he8}. We observe that the EIHH method seems to be less
effective for $^8$He than for $^6$He, because the second point in
$K_{\rm max}$ is still a few MeV away from the reference of the
extrapolation.  It is interesting to note that, even if not very
accurate, the extrapolated HH lies close to the coupled-cluster result
obtained in~\cite{Sonia_EPJ} where singles and doubles excitations
plus approximated triples were taken into account. As error for the
latter we take 10$\%$ of the difference with respect to the
Hartree-Fock calculation with the same potential, shown in
Fig.~\ref{he8} by the dotted line.

\subsection{Comparison with experiment}

\begin{figure}[t]
\begin{center}
\includegraphics[width=0.46\textwidth,clip=]{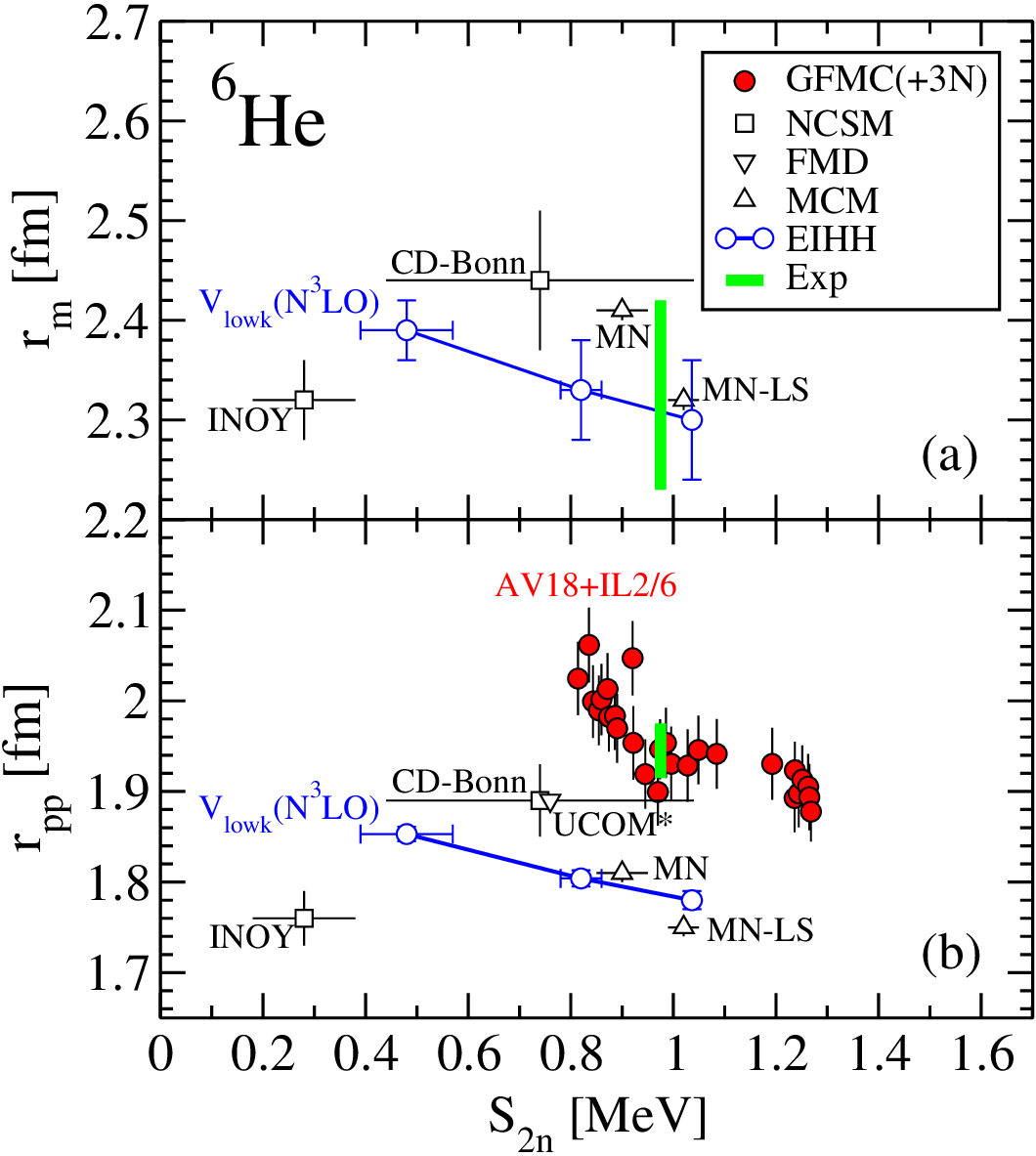}
\caption{Correlation plot of the $^6$He matter radius (upper panel)
  and point-proton radius (lower panel) versus two-neutron separation
  energy $S_{2n}$. The experimental range, shown by the bar (see text
  for details), is compared to theory based on different methods (NCSM~\cite{NCSMHe}, FMD~\cite{FMD},
  MCM~\cite{Brida} and our EIHH results using different NN
  interactions only and including 3N forces fit to light nuclei (only
  for GFMC~\cite{GFMC_Enrico_Fermi_School}). Calculational error bars
  are shown when available.
\label{fig:He6_correlation}}
\end{center}
\end{figure}

Next, we compare our results to the available experimental data and to
other {\it ab-initio} calculations.

The matter radius can be obtained from ion-scattering experiments,
which however is considerably uncertain. We include three different
data: in~\cite{Tan92} $r_{\rm m}$ was measured from a combined
analysis of the interaction cross section of $^6$He with a carbon
target and two-neutron removal cross sections of $^6$He projectiles,
leading to $2.33\pm0.04$ fm; in~\cite{Alk97} and~\cite{Kis05} $r_{\rm
  m}$ was measured from proton elastic scattering in inverse
kinematics leading to $2.30\pm0.07$ fm and $2.37\pm0.05$ fm,
respectively.  In Fig.~\ref{fig:He6_correlation}(a) we present these
data as a (green) band, which spans the three values with their
associated error bars.

The charge radius $r_{\rm ch}$ of halo nuclei, instead, can be
precisely and accurately measured via laser spectroscopy
techniques~\cite{Wan04,Mue07}.  The extraction of $r_{\rm ch}$ from
measured isotopic shifts requires very precise mass measurements and
atomic theory calculations.  The $^6$He charge radius has been
recently reevaluated using input from the first direct mass
measurement of this halo nucleus, leading to $r_{\rm
  ch}=2.060\pm0.008$ fm~\cite{Maxime}.  In order to compare the
experimental charge radius with theory, we convert it into a
point-proton radius $r_{\rm pp}$ using~\cite{Ong10}:
\begin{equation}
r^{2}_{\rm pp} = r^{2}_{\rm ch} - R^{2}_{p} - (N/Z) \cdot R^{2}_{n} -
3/(4 M^{2}_{p}) - r^{2}_{\rm so}\,,
\label{rpp}
\end{equation}
where $R^{2}_{p}$ and $R^{2}_{n} = -0.1161(22)$ fm$^{2}$ are the
proton and neutron mean-square charge radii, respectively, $3/(4
M^{2}_{p}) = 0.033$ fm$^{2}$ is a first-order relativistic
(Darwin-Foldy) correction~\cite{Fri97} and $r^{2}_{\rm so}$ is a
spin-orbit nuclear charge-density correction.  The latter should be
calculated from {\it ab-initio} wave functions and we will discuss our
results later. Because such a calculation is not available for all 
 methods, we prefer to use a common estimate for
$r^{2}_{\rm so}$ in the conversion of the experimental charge radius
to $r^2_{\rm pp}$.  In~\cite{Ong10}, the spin-orbit correction was
estimated to be $-0.08$ fm$^2$
in the case of pure $p_{3/2}$
halo neutrons for $^{6}$He.  
We conservatively took 0.08 fm$^2$
as the corresponding error.  For $R_{p}$ the Review of Particle
Physics~\cite{Nak10} value is $0.877(7)$ fm. Recently, $R_{p}$ has
been also precisely measured from spectroscopy of muonic
hydrogen~\cite{Poh10} leading to $0.84184(67)$ fm.  Using these two
values for $R_{p}$ with the above mentioned spin-orbit corrections in
Eq.~({\ref{rpp}}) we obtain $r_{\rm pp}=1.938\pm0.023$ fm and
$1.953\pm0.022$ fm for $^{6}$He, respectively.  The experimental
(green) band in Fig.~\ref{fig:He6_correlation}(b) includes both values
with their errors.

In order to present a combined comparison of our results to experiment
we show plots of the matter radius $r_{\rm m}$ and point-proton
$r_{\rm pp}$ radius versus the two-neutron separation energy $S_{2n}$.
The cutoff dependence of our EIHH results based on $\vlowk$ chiral
potentials allows us to study the correlation between these
observables, as shown in Fig.~\ref{fig:He6_correlation}.  We observe
that both the matter radius, in panel (a), and the point-proton
radius, in panel (b), increase when the separation energy decreases.
A smaller separation energy leads to a more extended halo structure
and thus larger $r_{\rm m}$ and $r_{\rm pp}$. Even though not
unexpected, it is interesting to see that such a correlation is
obtained from a set of phase-shift-equivalent interactions.  The lower
separation energy and larger radii are found for $\Lambda=2.4$
fm$^{-1}$. This indicates that for larger cutoff values the $^6$He
nucleus is unbound, as is the case with the Argonne $v_{18}$ (AV18)
potential~\cite{GFMC_6_8nuclei}.

The correlation band obtained from the EIHH results goes through the
experimental range for $r_{\rm m}$, which has a large uncertainty, but
does not go through the experimental range for $r_{\rm pp}$. Due to
the smaller uncertainty, this poses a stronger test for theory.
Before discussing more the comparison between theory and experiment,
we also show the results of other {\it ab-initio} calculations. The
GFMC energies~\cite{GFMC_Enrico_Fermi_School} are the only existing
converged calculations with 3N forces. Here, the employed
phenomenological potentials are constrained to reproduce the
properties of light nuclei, including $^6$He.
In~\cite{GFMC_Enrico_Fermi_School} it is explained that the GFMC
method does not reproduce the radii of halo nuclei as precisely as
energies and spectra of light nuclei, hence the different points in
Fig.~\ref{fig:He6_correlation}(b). The scatter in
Fig.~\ref{fig:He6_correlation}(b) gives some measure of the
uncertainty in the GFMC method as well as an uncertainty in the 3N
force models used [the IL2 and IL6 three-body forces were used with
the AV18 NN potential]~\cite{GFMC_Enrico_Fermi_School}.

Calculations with NN forces only include the FMD results based on the
UCOM potential plus a phenomenological correction to account for
three-body physics (which we denote with UCOM$^*$)~\cite{FMD}, the
NCSM results based on the CD-Bonn and INOY potentials~\cite{NCSMHe},
and variational MCM results based on the Minnesota (MN) and MN with
spin-orbit (MN-LS) potentials~\cite{Brida}.
Figure~\ref{fig:He6_correlation} shows that all theoretical results
based on NN forces only are compatible with the large experimental
range for $r_{\rm m}$, but not with $r_{\rm pp}$. They consistently
lie at lower $S_{2n}$ and smaller $r_{pp}$ values. The comparison to
theory in Fig.~\ref{fig:He6_correlation}(b) clearly highlights the
importance of including 3N forces. Figure~\ref{fig:He6_correlation}
also shows the importance of comparing theoretical predictions to more
than one observable. To illustrate this, both NCSM (using CD-Bonn) and
the GFMC results show a good agreement for the point-proton radius,
while the NCSM result has a large error for $S_{2n}$ and tends to
underpredict the two-neutron separation energy.

The observation that the $r_{\rm pp}-S_{2n}$ correlation band shown by
the EIHH results does not go through experiment is similar to the
Phillips and Tjon lines in few-body systems~\cite{few}, when only NN
interactions are included. Three-body physics manifests itself as a
breaking from this line/band. This behavior is also supported by the
variational MCM results.

Finally, we discuss the spin-orbit radius.  The $r^2_{\rm so}$ term is
a leading-order relativistic correction to the charge radius and
should be calculated from consistent {\it ab-initio} wave functions.
Following~\cite{Friar_Negele} and deriving the form of the 
$r^{2}_{\rm so}$ operator from the relativistic spin-orbit correction
to the charge density, we obtain
\begin{equation}
r^{2}_{\rm so}=\frac{1}{Z} \sum_i \frac{2\mu_i-e_i}{2m^2} \,
{\boldsymbol \sigma}_i \cdot {\boldsymbol \ell}'_i \,,
\label{rso}
\end{equation}
where $m, \mu_i$ and $e_i$ are the nucleon mass, magnetic moment and
charge, respectively. Here, ${\boldsymbol
\sigma}_i=2{\bf s}_i$ and ${\boldsymbol \ell}'_i$ represents the nucleon
angular momentum in the center-of-mass frame, defined as
\begin{equation}
{\boldsymbol \ell}'_i = ({\boldsymbol r}_i -  {\boldsymbol R}_{\rm cm})
\times \left({\boldsymbol p}_i  -\frac{1}{A} {\boldsymbol P}_{\rm cm}
\right) \,.
\label{ell}
\end{equation}
The estimate for the spin-orbit correction from
a shell-model picture based on pure 
%$P$-wave
$p_{3/2}$ halo neutrons is $-0.08$ fm$^2$
~\cite{Ong10}.  A microscopic calculation
based on the Minnesota potential gives
$-0.0718$ fm$^2$~\cite{Papadim}.
\begin{table}
\caption{$^6$He spin-orbit relativistic correction to the point-proton
  radius for the different cutoffs $\Lambda=1.8,2.0,2.4$ fm$^{-1}$ obtained
  from the EIHH method with $K_{\rm max}=10$.}
\begin{center}
\begin{tabular}{c|c}
\hline\hline
$\Lambda$ [fm$^{-1}$]&  $^6$He $\langle r^2_{\rm so}\rangle$ [fm$^2$]  \\
\hline
1.8 &  -0.0828\\
2.0 &  -0.0822\\
2.4 &  -0.0808\\
\hline\hline
\end{tabular}
\end{center}
\label{table_rso}
\end{table}

Using a harmonic-oscillator potential and $K_{\rm max}=2$ in the HH
expansion we can numerically reproduce the analytical results
of~\cite{Ong10}, if we correct for the center-of-mass as explained
below. Working with antisymmetrized wave functions, one has $\langle
\psi_0 \left| \sum_i O_i\right| \psi_0\rangle =A \langle \psi_0 \left|
O_A \right| \psi_0\rangle$.  For the operator in Eq.~(\ref{rso}), one
has to consider the spin-orbit term $ {\boldsymbol \sigma}_A \cdot
{\boldsymbol \ell}'_A$. In a shell-model picture one usually works
in the lab system so that the operator ${\boldsymbol \ell}'_A$ is
replaced with ${\boldsymbol \ell}_A={\boldsymbol r}_A\times
{\boldsymbol p}_A$. To see how ${\boldsymbol \ell}_A$ and
${\boldsymbol \ell}'_A$ are related, recall that in the coordinate
system of Eq.~(\ref{jacobc}) the following holds
\begin{eqnarray}
{\boldsymbol r}_A&=&\sqrt{\frac{A-1}{A}}{\boldsymbol
  \eta}_{A-1}+\frac{1}{\sqrt{A}}{\boldsymbol \eta}_0 \,, \\
{\boldsymbol p}_A&=&\sqrt{\frac{A-1}{A}}{\boldsymbol
  q}_{A-1}+\frac{1}{\sqrt{A}}{\boldsymbol q}_0 \,,
\end{eqnarray}
where ${\boldsymbol q}_{A-1}$ and ${\boldsymbol q}_{0}$ are the
conjugate coordinates of ${\boldsymbol \eta}_{A-1}$ and ${\boldsymbol
  \eta}_{0}$, respectively.  Taking the nucleus to be at rest,
${\boldsymbol q}_0=0$, one has
\begin{equation}
{\boldsymbol \ell}_{A}=\frac{A-1}{A}{\boldsymbol \ell}'_{\eta_{A-1}} + \sqrt{\frac{A-1}{A^3}} {\boldsymbol \eta}_0\times {\bf q}_{A-1}\,,
\end{equation}
where it can be shown that ${\boldsymbol
  \ell}_{\eta_{A-1}}={\boldsymbol \eta}_{A-1}\times {\boldsymbol
  q}_{A-1}={\boldsymbol \ell}'_A$.  Assuming that the wave function
factorizes in spherically symmetric center-of-mass and
intrinsic components, the term $\langle \phi_{\rm cm} | {\boldsymbol
  \eta}_0| \phi_{\rm cm}\rangle=0$, so that
\begin{equation}
{\boldsymbol \ell}'_A=\frac{A}{A-1} {\boldsymbol \ell}_A.
\end{equation}
Thus, in comparing to~\cite{Ong10} a factor of $6/5$ for $^6$He should
be included when intrinsic operators are used, leading to $\langle
r^2_{\rm so}\rangle=-0.096$ fm$^2$.
In Table \ref{table_rso}, we give results for $ \langle r^2_{\rm so}
\rangle = \langle \psi_0 \left| r^2_{\rm so}\right| \psi_0\rangle$
obtained with the EIHH method for the three different low-momentum
potentials employed. Calculations have been performed up to 
$K_{\rm max}=10$ which are converged at the few percent level.
Our  calculations deviate about 15$\%$ from the pure  $p_{3/2}$ halo-neutrons estimate.

We observe that the spin-orbit correction to the radius is small and
not enough to reconcile theory with experiment at the NN forces level.
When used in Eq.~(\ref{rpp}) to convert the experimental charge radius
to a point-proton radius, it leads to an enhancement of $r_{\rm pp}$,
going in the opposite direction with respect to what is needed to
improve the agreement with experiment.

\section{Conclusions}

We have carried out {\it ab-initio} calculations of the energy and
radii of $^6$He, using two-body $\vlowk$ interactions for different
cutoffs based on a chiral EFT potential~\cite{EM_pot}.  This work is
part of an ongoing effort to utilizing chiral EFT to predict the
properties of halo nuclei.  The calculations were performed using the
EIHH method.  The EIHH ground-state energies are in agreement with
previously published values obtained from extrapolating HH results.
The effective interaction greatly improves the convergence of the
radii of $^6$He, allowing us to avoid extrapolations.  The neutron
halo of $^6$He manifests itself via a 30$\%$ (40$\%$) enhancement of
the matter radius (point-neutron radius) with respect to the
point-proton radius.

In comparing with experiment, we have presented plots of the radii
versus the two-neutron separation energy.  We observe a correlation
between radii and $S_{2n}$ using the set of phase-shift-equivalent NN
potentials obtained by varying the resolution scale.  The correlation
band overlaps with the experimental data only for the matter radius,
which is relatively uncertain.  For the accurately measured charge
radius and inferred value of the point-proton radius, our results lie
at too low $r_{\rm pp}$ and $S_{2n}$ with respect to experiment. This
is a general trend for {\it ab-initio} results from NN interactions
only.  We have investigated whether the spin-orbit relativistic
correction to the radius, needed for the comparison between theory and
experiment, has an effect. From our  calculations we
observe that this correction is small and not enough to reconcile the
differences with experiment at the NN forces level. Comparing our
results to experiment and to the other {\it ab-initio} calculations it
is evident that 3N forces are crucial. Efforts to include chiral 3N
forces in our calculations are under way.

\section{Acknowledgment}

We thank M.\ Brodeur and J.\ Dilling for discussions. This work was
supported in part by the Natural Sciences and Engineering Research
Council (NSERC) and the National Research Council (NRC) of Canada, the
Israel Science Foundation (Grant No.~954/09), the Helmholtz Alliance
Program of the Helmholtz Association, contract HA216/EMMI ``Extremes
of Density and Temperature: Cosmic Matter in the Laboratory'', and the
DFG through grant SFB 634. Numerical calculations were performed at
TRIUMF.

\end{document}